\documentclass[aps,prX,preprint,superscriptaddress,onecolumn,nofootinbib]{revtex4-2}

\usepackage{amsmath,amssymb,amsfonts}
\usepackage{graphicx}
\usepackage{physics}
\usepackage{color}
\usepackage{hyperref}
\usepackage{bm}
\usepackage{setspace}

\hypersetup{
    colorlinks=true,
    linkcolor=blue,
    citecolor=blue,
    urlcolor=blue
}

\begin{document}

\title{Strong-Field Quantum Metrology Beyond the Standard Quantum Limit}

\author{Tsendsuren Khurelbaatar}
\email{t.khurelbaatar@griffith.edu.au}
\affiliation{Quantum and Advanced Technologies Research Institute, Griffith University, Brisbane, Queensland 4111, Australia}
\affiliation{School of Environment and Science, Griffith University, Brisbane, Queensland 4111, Australia}
\author{R.T. Sang}
\email{robert.sang@unisq.edu.au}
\affiliation{Quantum and Advanced Technologies Research Institute, Griffith University, Brisbane, Queensland 4111, Australia}
\affiliation{School of Environment and Science, Griffith University, Brisbane, Queensland 4111, Australia}
\affiliation{Office of the Pro Vice-Chancellor (Research, Development and Commercialisation), University of Southern Queensland, Springfield, 4300, Australia}
\author{Igor Litvinyuk}
\email{i.litvinyuk@griffith.edu.au}
\affiliation{Quantum and Advanced Technologies Research Institute, Griffith University, Brisbane, Queensland 4111, Australia}
\affiliation{School of Environment and Science, Griffith University, Brisbane, Queensland 4111, Australia}


\begin{abstract}
Bridging quantum optics and strong-field physics provides a pathway to explore how quantum light shapes extreme nonlinear light–matter interactions. However, direct characterization of non-classical light at damage-threshold intensities remains an open question. Here, we theoretically investigate the impact of photon-number fluctuations of squeezed light on strong-field photoelectron holography using a quantum-optical strong-field approximation. We identify a mechanism, \textit{ponderomotive dephasing}, whereby the inherent quantum fluctuations of the driving field dictate the stability of the electron's semiclassical action. While amplitude-squeezed light stabilizes the action to enhance holographic contrast, phase-squeezed light amplifies photon-number noise, causing a rapid collapse of fringe visibility. This quantum-optical sensitivity follows a steep quartic wavelength scaling ($\mathcal{V} \propto e^{-\lambda^4}$), rendering mid-infrared drivers uniquely sensitive to the field's underlying quantum nature.

Crucially, we show that the collapse of holographic contrast is not a loss of information but a metrological gain. By evaluating the Classical Fisher Information, we identify a ``dark-port" mechanism in the tunneling tail that enables the estimation of field quadrature noise beyond the Standard Quantum Limit. This fundamental trade-off between structural imaging fidelity and statistical sensitivity establishes the framework for \textit{Attosecond Quantum Tomography}: an \textit{in-situ}, reference-free protocol to reconstruct the Wigner distribution of intense quantum light. Our results identify strong-field ionization as a nonlinear quantum transducer, bridging attosecond electron dynamics with quantum information science.
\end{abstract}

\maketitle
\section{Introduction}

The semiclassical approximation, in which the atom is treated quantum mechanically while the driving field is described as a classical wave, has been the foundation of strong-field physics for several decades~\cite{corkum_plasma_1993,Lewenstein1994}. This framework relies on the assumption that the large photon occupation numbers ($\langle n \rangle \sim 10^{15}$) typical of processes such as high-harmonic generation render quantum vacuum fluctuations negligible. As experimental capabilities advance toward generating intense quantum-light sources~\cite{finger_raman-free_2015, perez_bright_2014,venneberg_bright_2025,sennary_attosecond_2025}, however, this assumption warrants closer examination. Recent theoretical studies indicate that non-classical photon statistics can leave measurable signatures in strong-field observables, including ionization yields and electron dynamics~\cite{rivera-dean_microscopic_2025, Gorlach2020, Gombkoto2020, Tzur2023, habibovic_intensity-dependent_2025, wang_attosecond_2025}, suggesting that aspects of the semiclassical description may require revision in this regime.
\begin{figure}[t]
    \centering
    \includegraphics[width=\linewidth]{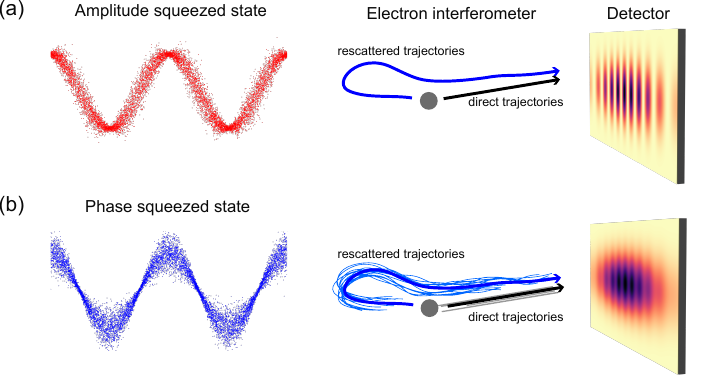}
    \caption{\textbf{Concept of ponderomotive dephasing.}
    \textbf{(a)} For an amplitude-squeezed driving field, photon-number fluctuations are
    suppressed, stabilizing the laser intensity and the ponderomotive energy $U_p$, and
    preserving high-contrast interference between direct and rescattered electron
    trajectories (blue and black arrows).
    \textbf{(b)} For a phase-squeezed field, the conjugate enhancement of photon-number
    fluctuations induces shot-to-shot variations of $U_p$. These intensity fluctuations
    lead to stochastic changes in the relative semiclassical action accumulated along the
    interfering trajectories, resulting in a collapse of the ensemble-averaged holographic
    interference contrast.}
    \label{fig:concept}
\end{figure}
A central challenge in exploring this interface is metrology. Characterizing non-classical light at intensities exceeding $10^{13}$~W/cm$^2$ is difficult using conventional quantum-optical tomography, which typically relies on combining the signal with a well-controlled local oscillator on optical components that cannot tolerate such field strengths and may alter inherent properties. Accessing quantum statistical properties under strong-field conditions therefore calls for detection mechanisms intrinsic to the interaction itself, effectively placing the measurement process \textit{inside} the laser focus.
Strong-Field Photoelectron Holography (SFPH)~\cite{bian_subcycle_2011,Huismans2011} provides a natural platform to explore this possibility. SFPH functions as a sub-cycle electron interferometer, encoding the phase difference between a ``reference'' wavepacket that reaches the detector directly and a ``signal'' wavepacket that undergoes rescattering from the parent ion (Fig.~\ref{fig:concept}). Traditionally, this technique has been employed to image atomic and molecular structure and electron dynamics~\cite{khurelbaatar_strong-field_2024, hasan_strong-field_2025}. Owing to the strong nonlinearity of the semiclassical action, however, the same interferometric phase is expected to be highly sensitive to fluctuations in the driving field, suggesting a pathway through which quantum optical noise could influence measurable photoelectron interference.

Here we explore this interface using a quantum-optical strong-field approximation (QO-SFA), revealing how the quantum statistics of intense light are directly imprinted onto photoelectron holography. In the tunneling regime, photon-number fluctuations of the driving field induce stochastic variations of the electron’s ponderomotive action, leading to a collapse of ensemble-averaged holographic contrast. We show that this apparent decoherence constitutes a metrological resource rather than a loss of information. By analyzing the Classical Fisher Information (CFI) of the photoelectron signal, we identify a regime in which strong-field ionization functions as a nonlinear transducer of quadrature-dependent quantum noise, enabling sensitivity beyond the standard quantum limit (SQL). This establishes a fundamental imaging-sensing duality in SFPH and provides the conceptual basis for attosecond-scale quantum tomography of intense non-classical light.

\section{Results}
\subsection{Theoretical Framework: The Quantum-Optical SFA}
To investigate the impact of quantum-light statistics on SFPH, we characterize the driving laser field using the Wigner phase-space representation~\cite{wigner_quantum_1932}.
A squeezed coherent state $|\alpha,\xi\rangle$ is defined by a complex displacement
$\alpha=|\alpha|e^{i\phi}$ and a squeezing parameter $\xi=re^{i\theta}$, where $r$ denotes
the squeezing magnitude and $\theta$ the squeezing angle.

In our model, the peak field amplitude $E_0$ is treated as a stochastic variable sampled
from the Wigner distribution $W(\alpha,r,\theta)$:
\begin{equation}
W(\mathbf{X}) =
\frac{1}{2\pi\sqrt{\det\boldsymbol{\sigma}}}
\exp\!\left[
-\frac{1}{2}(\mathbf{X}-\bar{\mathbf{X}})^T
\boldsymbol{\sigma}^{-1}
(\mathbf{X}-\bar{\mathbf{X}})
\right],
\end{equation}
where $\mathbf{X}=(X_1,X_2)$ denotes the field quadratures.
Here $\boldsymbol{\sigma}$ is the $2\times2$ covariance matrix,
$\sigma_{ij}=\langle\Delta X_i \Delta X_j\rangle$,
which for a squeezed coherent state is fully determined by the squeezing magnitude $r$
and angle $\theta$, and $\bar{\mathbf{X}}=\langle\mathbf{X}\rangle$ denotes the phase-space
displacement corresponding to the coherent amplitude $\alpha$.

For the numerical simulations, we consider atomic hydrogen driven by a linearly polarized
Gaussian laser pulse with a central wavelength of 1.5\,$\mu$m and a peak intensity of
$1\times10^{14}\,\mathrm{W/cm^2}$. To isolate intracycle holographic interference while
suppressing intercycle contributions, the electron dynamics are restricted to a single
optical cycle. Field fluctuations are incorporated using the phase-space QO-SFA, in which the driving field is described by its Wigner
quasiprobability distribution $W(\alpha)$~\cite{Glauber1963,Cahill1969}. Each realization $\alpha$
defines the peak field amplitude of the sampled field realization, ensuring that the quantum statistics
of the source are consistently mapped onto sub-cycle electron trajectories.

The photoelectron momentum distribution (PMD) $P(\mathbf{p})$ is obtained as an incoherent
ensemble average over single-shot transition amplitudes $M(\mathbf{p};\alpha)$, evaluated
within the standard SFA saddle-point formalism~\cite{kopold_quantum_2000,milosevic_role_2002}:
\begin{equation}
P(\mathbf{p}) = \int d^2\alpha \, W(\alpha) \, \left| M(\mathbf{p}; \alpha) \right|^2 .
\label{eq:ensemble_avg}
\end{equation}
This formulation establishes a direct and quantitative link between non-classical light and strong-field electron observables.

\subsection{Quantum-State Control of Photoelectron Holography}
\begin{figure*}[t]
    \centering
    \includegraphics[width=\linewidth]{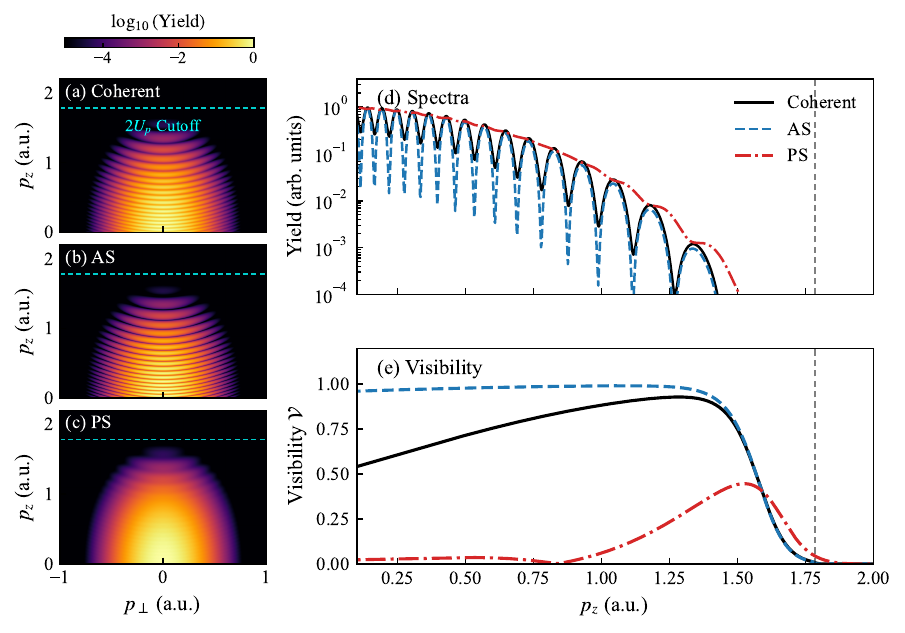}
    \caption{\textbf{Quantum-state control of photoelectron holography.} 
        \textbf{(a--c)} Photoelectron momentum distributions (PMDs) driven by a 1.5\,$\mu$m field in three distinct quantum states: \textbf{(a)} Coherent state (CS), defining the standard quantum limit for holographic contrast; \textbf{(b)} Amplitude-squeezed (AS) state ($r=1.5$), where suppressed intensity fluctuations stabilize the electron action; and \textbf{(c)} Phase-squeezed (PS) state ($r=1.5$), where conjugate anti-squeezing of the photon number washes out interference. 
        \textbf{(d)} Photoelectron energy spectra along the polarization axis ($p_{\perp} = 0$). The PS state (red) exhibits a complete collapse of modulation depth compared to the deep holographic fringes of the AS state (blue dashed). 
        \textbf{(e)} Trajectory-resolved holographic fringe visibility $\mathcal{V}$ as a function of longitudinal momentum $p_z$. While the CS (black) maintains moderate visibility, the AS state achieves near-perfect coherence ($\mathcal{V} \approx 1$). In contrast, the PS state induces rapid dephasing, demonstrating that strong-field coherence is governed by photon-number statistics rather than optical phase stability.}
        \label{fig:pmd}
\end{figure*}
We now demonstrate that the quantum-light states provide a powerful control knob for intracycle photoelectron holography. The PMDs in Fig.~\ref{fig:pmd} reveal a pronounced dependence on the non-classical statistics of the driver, showing how quantum fluctuations of the driver are directly transduced into macroscopic strong-field ionization observables. The coherent-state (CS) driver (Fig.~\ref{fig:pmd}(a)) produces the well-known intracycle photoelectron holographic interference pattern~\cite{Faria2020}, which we take as the SQL for the present strong-field observable. Replacing the coherent state with an amplitude-squeezed (AS) state (Fig.~\ref{fig:pmd}(b)) preserves the holographic structure while significantly enhancing the fringe contrast. In stark contrast, driving with a phase-squeezed (PS) state (Fig.~\ref{fig:pmd}(c)) leads to a complete wash-out of the interference, resulting in a nearly featureless PMD.

This loss of coherence is quantified in Figs.~\ref{fig:pmd}(d,e). The photoelectron energy spectra along the polarization axis (Fig.~\ref{fig:pmd}(d)) show pronounced modulations for both the CS and AS cases, with the AS state exhibiting substantially deeper fringes, indicative of enhanced holographic contrast. By contrast, the PS spectrum displays only weak residual modulation and a largely monotonic decay. These trends are captured quantitatively by the trajectory-resolved fringe visibility $\mathcal{V}(p_z)$ (see Supplemental Material (SM)~\cite{SM}), shown in Fig.~\ref{fig:pmd}(e). For the AS state, near-unity visibility ($\mathcal{V}\approx1$) is maintained across the entire momentum range. In the CS case (black solid line), the visibility is reduced at low longitudinal momenta ($p_z<0.5$~a.u.) but increases steadily toward unity near the $2U_p$ cutoff.

This momentum dependence reflects the role of ponderomotive dephasing. The phase sensitivity to intensity fluctuations scales with the electron excursion time, $\tau_{\mathrm{exc}}$, in the continuum (see Sec.~\ref{sec:ponderomotive}). Low-momentum electrons originate from long trajectories with large $\tau_{\mathrm{exc}}$ accumulating substantial random phase noise ($\sigma_\phi^2\propto\tau_{\mathrm{exc}}^2$). Near the $2U_p$ cutoff, however, electrons arise from short trajectories with minimal excursion time, rendering the interference naturally more robust and leading to a recovery of visibility.

Most notably, the PS state exhibits a pronounced collapse of visibility for $p_z<1.0$~a.u., highlighting a qualitative departure from intuition based on linear interferometry, where phase stability is paramount~\cite{aasi_enhanced_2013}. In the strong-field regime, the stability of holographic interference is governed not by the optical phase of the driving field, but by its photon-number statistics.
These results demonstrate that, in SFPH, coherence is controlled primarily by photon-number fluctuations rather than optical phase stability.

\subsection{The Physical Mechanism: Ponderomotive Dephasing}
\label{sec:ponderomotive}

The squeezing-dependent visibility observed in Fig.~\ref{fig:pmd} originates from the intensity dependence of the electron’s semiclassical action in SFPH. In this framework, the interference pattern is encoded in the phase difference $\Delta S$ between a ``reference'' (direct) wavepacket and a ``signal'' (rescattered) wavepacket (Fig.~\ref{fig:concept})~\cite{Huismans2011,bian_subcycle_2011}. During the electron’s excursion in the continuum, this phase difference is dominated by the quiver motion in the laser field~\cite{shvetsov-shilovski_effects_2018} and can be approximated as
\begin{equation}
\Delta S(\mathbf{p}) \;\approx\; \alpha_0\, U_p\, \tau_{\mathrm{exc}} + C,
\label{eq:action_diff}
\end{equation}
where $\tau_{\mathrm{exc}}$ is the excursion time, $\alpha_0$ is a trajectory-dependent coefficient, and $C$ collects sub-dominant contributions from the ionization potential $I_p$ and the drift momentum $\mathbf{p}$. In the mid-infrared regime ($\lambda=1.5\,\mu$m), the ponderomotive energy $U_p$ dominates the action, allowing these additional terms to be treated as a constant offset when establishing scaling laws (see SM~\cite{SM}).
\begin{figure*}[t] 
    \centering
    \includegraphics[width=\linewidth]{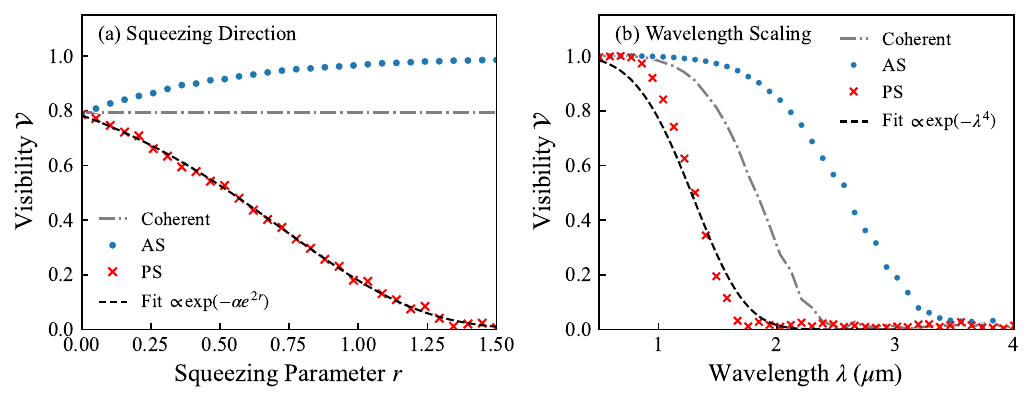}
    \caption{\textbf{Scaling laws of ponderomotive dephasing.} 
    (a) Holographic fringe visibility $\mathcal{V}$ as a function of the squeezing parameter $r$ at $\lambda=1.5\,\mu$m and fixed longitudinal momentum $p_z=0.8$~a.u. PS light (red crosses) exhibits a double-exponential decay, $\mathcal{V}(r)\propto\exp(-\eta e^{2r})$ (black dashed fit), driven by amplified photon-number fluctuations. In contrast, AS light (blue circles) suppresses ponderomotive shot noise, driving the visibility toward unity relative to the coherent-state baseline (gray dash-dotted line), which defines the SQL.
    (b) Visibility as a function of the driving wavelength $\lambda$ for $r=1$ and $p_z=0.8$~a.u. The PS state exhibits a pronounced ``quantum cliff'' in the mid-infrared, consistent with the predicted scaling $\mathcal{V}(\lambda)\propto\exp(-\beta\lambda^4)$. By contrast, the AS state remains robust across the spectrum, confirming that the observed dephasing originates from anti-squeezing of the ponderomotive quadrature.}
    \label{fig:scaling}
\end{figure*}
In a quantum-optical description, the ponderomotive energy becomes an operator proportional to the photon number, $\hat{U}_p \propto \hat{n}$, such that fluctuations of the field intensity translate directly into fluctuations of the electron action. This linear mapping couples the stability of the holographic phase to the photon-number statistics of the driving field. For a squeezed coherent state $|\alpha, r\rangle$ with $|\alpha|^2 \gg 1$, the Heisenberg uncertainty principle enforces a trade-off between phase and photon-number fluctuations. In particular, phase squeezing ($\Delta\phi \rightarrow 0$) necessarily induces anti-squeezing of the photon-number variance, leading to
\begin{equation}
\sigma_{U_p}^2 \;\propto\; \langle(\Delta\hat{n})^2\rangle \;\approx\; |\alpha|^2 e^{2r},
\label{eq:variance_up}
\end{equation}
These shot-to-shot fluctuations of the ponderomotive energy generate stochastic phase shifts between interfering trajectories, scrambling the holographic phase relation. We refer to this mechanism as \emph{ponderomotive dephasing}.

Assuming Gaussian photon-number statistics, the ensemble-averaged interference term obeys $\langle e^{i\delta\phi}\rangle = \exp[-\langle\delta\phi^2\rangle/2]$ for a zero-mean phase fluctuation $\delta\phi$. Identifying $\delta\phi \approx \kappa\,\delta U_p$, the holographic fringe visibility decays as (see SM~\cite{SM})
\begin{equation}
\mathcal{V}
= \left|\langle e^{i\kappa\,\delta U_p}\rangle\right|
= \exp\!\left(-\tfrac{1}{2}\kappa^2\sigma_{U_p}^2\right),
\label{eq:visibility_decay}
\end{equation}
where $\kappa$ is a trajectory-dependent sensitivity factor. Equation~\eqref{eq:visibility_decay} represents a leading-order result in the tunneling regime; Coulomb phase corrections are neglected but do not alter the scaling.

The consequences of this mechanism are quantified in Fig.~\ref{fig:scaling}(a), which reveals a clear bifurcation in holographic visibility as a function of the squeezing parameter $r$. For the AS state, photon-number fluctuations are suppressed below the shot-noise level, stabilizing the ponderomotive phase and driving the visibility toward unity. In contrast, PS state amplifies photon-number fluctuations, producing a rapid collapse of interference. The simulated data (red crosses) follow a double-exponential decay, $\mathcal{V}(r)\propto\exp(-\eta e^{2r})$, in excellent agreement with the analytical prediction of Eq.~\eqref{eq:visibility_decay}. This behavior confirms that, in SFPH, holographic coherence is governed not by the optical phase of the driving field, but by the quantum fluctuations of the field amplitude which dictates the stability of the semiclassical action.

\subsection{Wavelength scaling}

The susceptibility of the holographic pattern to ponderomotive dephasing increases strongly with the driving wavelength $\lambda$. To understand this scaling, we consider the electron’s accumulated phase variance, $\sigma_\phi^2$. For a fixed classical intensity ($I \propto \langle \hat{n} \rangle / \lambda$), the electron excursion time scales linearly with wavelength, $\tau_{\mathrm{exc}} \propto \lambda$. In the quantum-optical description, the ponderomotive energy operator scales as $\hat{U}_p \propto \lambda^2 (\hat{n}/\lambda) \propto \lambda \hat{n}$.

Combining these dependencies, the phase fluctuations scale as $\sigma_\phi \sim \tau_{\mathrm{exc}} \sigma_{U_p} \propto \lambda^2 \sigma_n$, leading to a steep quartic scaling of holographic visibility:
\begin{equation}
\ln \mathcal{V} \propto -\sigma_\phi^2 \propto -\lambda^4 \langle (\Delta \hat{n})^2 \rangle.
\label{eq:quartic_scaling}
\end{equation}
This scaling has no analogue in linear interferometry and arises from the interplay between the wavelength-dependent excursion time and ponderomotive energy fluctuations.

Figure~\ref{fig:scaling}(b) illustrates this behavior. The CS baseline (gray dash-dotted line) remains relatively stable up to $\sim1.5~\mu$m, while PS light (red crosses) exhibits a sharp drop in visibility beyond this wavelength, forming the pronounced ``quantum cliff'' predicted by Eq.~\eqref{eq:quartic_scaling}. The black dashed fit confirms the quartic scaling, $\ln \mathcal{V} \propto -\lambda^4$, highlighting that mid-infrared drivers amplify quantum fluctuations. In contrast, AS light (blue circles) remains robust up to $\sim2~\mu$m, demonstrating that the dephasing is an asymmetric effect arising specifically from photon-number fluctuations.

These results imply that, as attosecond science progresses toward long-wavelength drivers, the quantum statistics of the field will become the dominant factor in determining holographic contrast. Consequently, standard semiclassical treatments may significantly underestimate the role of quantum noise, emphasizing the necessity of a full quantum-optical treatment.

\subsection{The metrological duality: trade-off between imaging and sensing}
The collapse of holographic visibility observed in Fig.~\ref{fig:pmd}(c) does not imply a loss of information; rather, it signals a transition in the role of the strong-field interaction from imaging to sensing. In this regime, SFPH functions as a quadrature-sensitive, reference-free measurement of the driving field, in which quantum fluctuations are transduced into measurable variations of the photoelectron signal.

To quantify this transition, we evaluate the CFI, $\mathcal{F}$, which sets the lower bound on the precision of parameter estimation via the Cram\'er--Rao inequality~\cite{caves_quantum-mechanical_1981,giovannetti_advances_2011}. For coherent light, the CFI defines the SQL operationally as the estimation performance obtained with a coherent-state driver for the same photoelectron observable, providing a natural baseline for comparison. Introducing squeezing modifies the photon-number statistics of the driving field and leads to a monotonic enhancement of the CFI relative to this baseline.
\begin{figure}[t]
    \centering
    \includegraphics[width=\linewidth]{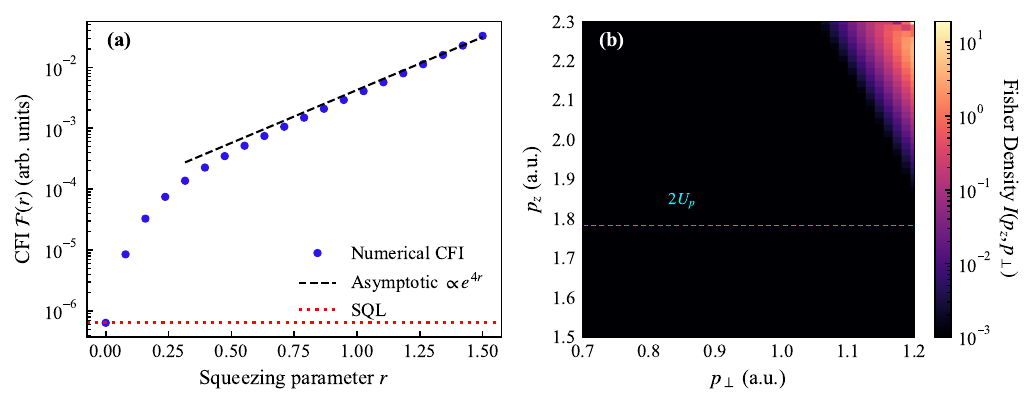}
    \caption{\textbf{Metrological performance.} 
    (a) Integrated Classical Fisher Information (CFI) as a function of the squeezing parameter $r$ for the PS state. Numerical results (blue circles) show a pronounced enhancement over the SQL (red dotted line). In the high-squeezing regime ($r>0.5$), the CFI follows an asymptotic scaling $\mathcal{F}\propto e^{4r}$ (black dashed line), reflecting the nonlinear amplification of quantum fluctuations in the driving field by the atomic response. 
    (b) Two-dimensional Fisher information density $I(p_z,p_\perp)$ in the high-momentum cutoff region. The dashed cyan line indicates the classical cutoff energy ($2U_p$). The information is concentrated in the deep tunneling tail ($p_z>2U_p$), exploiting a ``dark port'' mechanism in which the exponentially suppressed electron yield provides maximal relative sensitivity against vacuum fluctuations.}
    \label{fig:fisher_scaling}
\end{figure}
Strong-field ionization is a highly nonlinear, non-unitary process in which the measured observable, the PMD, depends exponentially on the field amplitude through the semiclassical action. The atom thus acts as a nonlinear quantum transducer that maps photon-number fluctuations of the driving field onto macroscopic variations of the electron yield, rendering Quantum Fisher Information bounds derived for linear optical interferometry inapplicable~\cite{giovannetti_quantum-enhanced_2004}.

Figure~\ref{fig:fisher_scaling}(a) shows that the resulting CFI increases rapidly with the squeezing parameter $r$, demonstrating a pronounced metrological gain enabled by the nonlinear strong-field response. The physical origin of this enhancement is clarified by the two-dimensional Fisher information density shown in Fig.~\ref{fig:fisher_scaling}(b). The information is concentrated in a ``dark-port'' region beyond the classical $2U_p$ cutoff, where the electron yield is exponentially suppressed while its relative sensitivity to field fluctuations is maximized. This mechanism is analogous to homodyne detection at the destructive interference port~\cite{caves_quantum-mechanical_1981}; the same vacuum fluctuations that scramble holographic fringes in the plateau region become the primary resource for precision estimation in the tunneling tail.

These results establish a metrological duality intrinsic to SFPH. The AS state suppresses photon-number fluctuations, stabilizing the ponderomotive phase and enabling high-fidelity holographic imaging. In contrast, the PS state amplifies amplitude noise, leading to rapid visibility decay but simultaneously maximizing sensitivity in the dark-port cutoff region. Here, the loss of interference contrast is not detrimental; instead, it constitutes a macroscopic signature of enhanced metrological power, allowing SFPH to serve as a self-referenced probe of the quantum statistics of intense quantum light.

\section{Discussion}

The results presented here place SFPH in a qualitatively new regime: one in which the quantum statistics of the driving field act as an intrinsic control parameter of the electron interferometer. In conventional SFPH experiments, loss of holographic contrast is typically attributed to technical decoherence, such as focal averaging or classical intensity noise. However, our quantum-optical treatment shows that, beyond the classical approximation, ponderomotive dephasing exhibits a steep wavelength dependence, giving rise to a ``quantum cliff'' that naturally explains the long-standing observation that holographic contrast often degrades more rapidly at long wavelengths than predicted by semiclassical models. This mechanism is not specific to squeezed light, but represents an amplified manifestation of a general limit that becomes unavoidable with mid-infrared drivers. 

A key insight is that strong-field electron interference is governed not by optical phase stability, as in linear interferometry, but by fluctuations of the ponderomotive energy accumulated during the electron’s excursion in the continuum. The PS states, which enhance sensitivity in conventional interferometers, induce rapid dephasing in SFPH, whereas AS states stabilize the semiclassical action and preserve high-contrast interference. The electron interferometer thus selects an intrinsic optical quadrature determined by the nonlinear mapping from photon number to semiclassical action, with no analogue in linear optics. While extensions incorporating Coulomb phases, multielectron dynamics, or propagation effects may refine quantitative predictions, they are unlikely to alter the core qualitative insights.

The experimental observation of ponderomotive dephasing and the quantum cliff is within reach of current state-of-the-art squeezed-light sources. In our simulations, the transition from the standard quantum limit to the high-sensitivity regime is captured using a squeezing parameter \(r\) ranging from 0.5 to 1.5. Measurable deviations in holographic contrast and CFI enhancement begin to emerge at \(r \approx 0.5\), corresponding to approximately 4.3 dB of squeezing. The pronounced collapse of visibility at 1.5\,$\mu$m, which marks the quantum cliff at \(r = 1.0\), requires 8.7 dB of squeezing, a level routinely achieved in modern parametric down-conversion and four-wave mixing experiments. Near-perfect fringe stabilization under amplitude squeezing and the \(e^{4r}\) asymptotic scaling of the CFI under phase squeezing, demonstrated at \(r = 1.5\), correspond to 13 dB of squeezing. While technically demanding at high intensities, recent advances in noble-gas-filled photonic-crystal fibers and nonclassical power stabilization show the feasibility of generating bright squeezed vacuum at these levels. Moreover, the use of mid-infrared drivers (e.g., 1.5–2.0\,$\mu$m) exploits the \(\lambda^4\) scaling of ponderomotive dephasing, effectively lowering the ``squeezing threshold'' required to distinguish quantum statistical effects from classical noise and making these effects experimentally observable even with moderate squeezing.

\section{Conclusion}

We have revealed a fundamental complementarity in strong-field physics that emerges at the quantum-optical limit: structural imaging fidelity and quantum-light sensitivity are mutually exclusive resources, constrained by the Heisenberg uncertainty principle. In the high-fidelity imaging regime, the atom acts as a sub-\AA{}ngstrom diffractometer, enabling structural imaging below the shot-noise limit of standard coherent fields. Under phase squeezing, the atom functions as a quantum-enhanced interferometer, where the collapse of visibility and the ``dark-port'' dynamics in the tunneling tail enable \textit{in situ} characterization of intense quantum light statistics beyond the SQL.

This complementarity highlights a conceptual shift in the understanding of strong-field interactions. By tailoring the photon statistics of high-power lasers, one can toggle the atomic response between structural imaging and quantum sensing, accessing information inaccessible to classical-field treatments.

Taken together, these results suggest a new perspective in ultrafast science, where strong-field interactions are not merely classical electron dynamics but also a platform for quantum-optical control. By exploiting the interplay between electron trajectories and the quantum statistics of the driving field, future experiments can achieve quantum-enhanced measurement and coherent control at extreme intensities. This framework demonstrates that even at the highest intensities, light–matter interactions remain fundamentally quantum in nature.


\bibliography{references}

\end{document}